\documentclass[usegraphicx]{mn2e}
\usepackage{subfig}
\usepackage{epsfig}
\usepackage{amsmath}
\usepackage{amssymb}
\usepackage{graphicx}
\usepackage{times,longtable,color}

\usepackage{wrapfig}

\newcommand {\apgt} {\ {\raise-.5ex\hbox{$\buildrel>\over\sim$}}\ }
\newcommand {\aplt} {\ {\raise-.5ex\hbox{$\buildrel<\over\sim$}}\ } 
\newcommand {\degree}{$^{\circ}$}

\title[On the magnetic field in M51 ULX-8]
{On the magnetic field in M51 ULX-8}

\author[M. Middleton et al.]
{M. J. Middleton$^{1}$, M. Brightman$^{2}$, F. Pintore$^{3}$, M. Bachetti$^{4}$, A. C. Fabian$^{5}$,  \newauthor  F. F{\"u}rst$^{6}$ \& D. J. Walton$^{5}$\\
\\
1. Department of Physics and Astronomy, University of Southampton, Highfield, Southampton SO17 1BJ, UK\\
2. Cahill Center for Astrophysics, California Institute of Technology, 1216 East California Boulevard, Pasadena, CA 91125, USA\\
3. INAF - IASF Milano, Via E. Bassini 15, I-20133 Milano, Italy\\
4. INAF-Osservatorio Astronomico di Cagliari, via della Scienza 5, I-09047 Selargius, Italy\\
5. Institute of Astronomy, University of Cambridge, Madingley Road, Cambridge CB3 0HA, UK\\
6. European Space Astronomy Centre (ESAC), Science Operations Departement, 28692 Villanueva de la Canada, Madrid, Spain\\
}

\pagerange{\pageref{firstpage}--\pageref{lastpage}} \pubyear{2017}
\long\def\symbolfootnote[#1]#2{\begingroup\def\thefootnote{\fnsymbol{footnote}}\footnote[#1]{#2}\endgroup} 
\def\ga{\mathrel{\hbox{\rlap{\hbox{\lower4pt\hbox{$\sim$}}}{\raise2pt\hbox{$>$}}
}}}

\begin{document}

\topmargin = -0.5cm

\maketitle

\label{firstpage}

\begin{abstract}

The reported discovery of a cyclotron resonance scattering feature (CRSF) in the spectrum of M51 ULX-8 may provide an important clue as to the nature of the magnetic field in those ultraluminous X-ray sources hosting neutron stars. In this paper we present the covariance (linearly correlated variability) spectrum of M51 ULX-8 on long ($>$ 2000s) timescales. This allows us to unambiguously decompose the spectrum which requires multiple components in order to explain the broad-band emission. Having a clearer picture of the spectral decomposition leads to various tests for the dipole field strength of the neutron star which can be extended to other ULXs when certain criteria are met. In the case of M51 ULX-8, we rule out a very strong ($\sim$10$^{15}$~G) dipole solution with either a sub- or super-critical disc. Instead, our tests indicate an upper limit on the dipole field of $\sim$10$^{12}$~G and a classical super-critical inflow, similar to that inferred in other ULXs found to harbour neutron stars, although we do not rule out the presence of an additional, strong ($\sim$10$^{15}$~G) multipole field falling off steeply with distance from the neutron star.

\end{abstract}


\section{introduction}

The discovery of ultraluminous pulsars (ULPs or PULXs, Bachetti et al. 2014; F{\"u}rst et al. 2016; Israel et al. 2017a, b; Carpano et al. 2018) has revolutionised the field of ultraluminous X-ray sources (ULXs, see the review of Kaaret et al. 2017). Whilst ULXs were long considered to be possible candidates for hosting intermediate mass black holes (IMBHs), it was immediately apparent that the explanation for the extreme observed luminosities ($>$ 10$^{39}$ erg~s$^{-1}$) in at least {\it some} ULXs was accretion in excess of the classical Eddington limit onto common-place primary objects --- in this case neutron stars. However, whilst the mass-regime of the compact object in ULXs has been at least partly resolved (we note that candidate IMBHs still remain, e.g. Farrell et al. 2009), the relative number of black hole to neutron star primaries in ULXs (see King, Lasota, \& Kluzniak 2017; Middleton \& King 2017) and the nature of the accretion flow in ULPs remain outstanding puzzles. At the centre of the debate is the strength of the surface dipole field and any multipolar component. Should the dipole field strength be similar to that of Galactic HMXBs ($\sim$10$^{12}$~G --- e.g. F{\"u}rst et al. 2014; Tendulkar et al. 2014; Yamamoto et al. 2014; Bellm et al. 2014) then it is quite plausible that the flow will be super-critical ($\dot{m}/\dot{m}_{\rm Edd} >$ 1 where $\dot{m}_{\rm Edd}$ is the Eddington accretion rate) at radii greater than the magnetospheric truncation radius ($r_{\rm M}$). In this case, the super-critical portion of the disc will have a large --- close to unity --- vertical scale-height and winds will be launched from the surface (see Shakura \& Sunyaev 1973; Poutanen et al. 2007 and the simulations of Ohsuga et al 2009; Sadowski et al. 2014; Jiang, Stone \& Davis 2014). Within $r_{\rm M}$, the flow will take the form of an accretion curtain (Mushtukov et al. 2017; 2019) and shock-heated column as material falls onto the magnetic poles. Due to collimation by the disc and outflows beyond $r_{\rm M}$, it is expected that the intrinsic luminosity is then partially geometrically beamed (see King 2009). Conversely, should the dipole field strength be very high (typically $>$ 10$^{13}$~G) then it is quite probable that the disc will truncate before becoming locally super-critical. The geometry in this case is then expected to take the form of a geometrically thin disc down to $r_{\rm M}$, an accretion curtain and shock-heated column. Rather than a super-critical disc and geometrical beaming, super-Eddington luminosities can then be explained by a magnetic pressure supported accretion column (Basko \& Sunyaev 1976) and high field strength, the latter allowing for a substantially increased luminosity from a reduction in the electron scattering cross-section (e.g. Herold 1979; Paczynski 1992; Thompson \& Duncan 1995; Mushtukov et al. 2015) 

\begin{table*}
\begin{center}
\begin{minipage}{140mm}
\bigskip
\caption{Spectral fitting results}
\begin{tabular}{|c|c|c|c|c|c}

\hline

\multicolumn{2}{c} {Covariance spectral fit} & \multicolumn{2}{c}{Time-averaged spectral fit (fixed line)} & \multicolumn{2}{c}{Time-averaged spectral fit (free line)}\\
\multicolumn{2}{c} {({\sc tbabs*gabs*cutoffpl})} & \multicolumn{2}{c}{({\sc tbabs*[diskbb + gabs*cutoffpl]})} & \multicolumn{2}{c}{({\sc tbabs*[diskbb + gabs*cutoffpl]})} \\
   \hline
   \hline

nH ($\times$10$^{22}$cm$^{-2})$ & 0.08 (fixed)  & nH (10$^{22}$cm$^{-2})$ &  0.05 $\pm$ 0.02 & nH (10$^{22}$cm$^{-2})$ &  0.11 $^{+0.04}_{-0.03}$\\
& & kT$_{\rm dbb}$ (keV) & 0.58 $\pm 0.03$ & kT$_{\rm dbb}$ (keV) & 0.41$^{+0.07}_{-0.08}$\\
& & N$_{\rm dbb}$ & 0.12 $_{-0.02}^{+0.03}$ & N$_{\rm dbb}$ & 0.46 $_{-0.22}^{+0.76}$\\
& & L$_{\rm dbb}$ ($\times$ 10$^{39}$ erg s$^{-1}$) & 2.33$^{+0.08}_{-0.07}$ & L$_{\rm dbb}$ ($\times$ 10$^{39}$ erg s$^{-1}$) & 2.03$^{+0.12}_{-0.14}$\\
$\Gamma$ & -0.75 $_{-0.81}^{+1.10}$  &  $\Gamma$ & -0.75 (fixed) &  $\Gamma$ & -0.75 (fixed)\\ 
E$_{\rm cutoff}$ (keV) & 1.61 $_{-0.40}^{+9.04}$  & E$_{\rm cutoff}$ (keV) & 1.61 (fixed) & E$_{\rm cutoff}$ (keV) & 1.61 (fixed)\\ 
N$_{\rm cutoff}$ ($\times$ 10$^{-5}$) & 2.05 $_{-0.77}^{+0.47}$  & N$_{\rm cutoff}$ ($\times$ 10$^{-5}$)  & 3.34 $_{-0.18}^{+0.17}$ & N$_{\rm cutoff}$ ($\times$ 10$^{-5}$)  & 5.30 $_{-0.85}^{+1.76}$ \\ 
& & L$_{\rm cutoff}$ ($\times$ 10$^{39}$ erg s$^{-1}$) & 2.52$^{+0.08}_{-0.09}$ & L$_{\rm cutoff}$ ($\times$ 10$^{39}$ erg s$^{-1}$) & 4.00$^{+0.31}_{-0.24}$\\
E$_{\rm cyc}$ (keV) & 4.52 (fixed) & E$_{\rm cyc}$ (keV) & 4.52 (fixed) & E$_{\rm cyc}$ (keV) & 4.72 $_{-0.13}^{+0.17}$\\
$\sigma$ (keV) & 0.11 (fixed) & $\sigma$ (keV) & 0.11 (fixed) & $\sigma$ (keV) & 0.96 $_{-0.26}^{+0.41}$\\
N$_{\rm cyc}$ & $<$ 263  & N$_{\rm cyc}$ & 0.23 $_{-0.05}^{+0.06}$ & N$_{\rm cyc}$   & 1.63 $_{-0.67}^{+1.57}$\\
$\chi^{2}$/d.o.f   & 1.16/2  & $\chi^{2}$/d.o.f   &  244.48/216 & $\chi^{2}$/d.o.f  & 239.56/214\\
null P value & 0.56 & null P value & 0.09 & null P value & 0.11 \\

\hline
\end{tabular}
Notes: Best-fitting model parameters for the fit to the covariance and time-averaged data of M51 ULX-8 (see Figure 2) including the unabsorbed 0.3-8~keV luminosity of each component. Formal errors are quoted at 1$\sigma$ (see Section 2 for further discussion).
\end{minipage} 
\end{center}
\end{table*}

Certainly there are compelling reasons to believe either scenario described above: the measured rate of spin-up for those ULPs found thus far would seem to imply dipole field strengths  $\sim$10$^{11}$ -- 10$^{13}$~G (e.g. King \& Lasota 2016; Christodoulou et al. 2016; F{\"u}rst et al. 2016;  King, Lasota \& Kluzniak 2017; Carpano et al. 2018), whilst a higher-strength dipole scenario can more easily explain the sinusoidal profile of the pulsations (Mushtukov et al. 2017). 

Reported in Liu \& Mirabel (2005), M51 ULX-8 is one of the brightest ULXs associated with this interacting galaxy and can be excluded as a foreground source from its identified stellar counterpart (Terashima et al 2006) and as an AGN from the strong curvature in the X-ray spectrum. With the identification of a cyclotron resonance scattering feature (CRSF) in M51 ULX-8 (Brightman et al. 2018) we have been provided an opportunity to directly probe the nature of the magnetic field in the vicinity of what is assumed to be a neutron star. As Brightman et al. (2018) report, if the line at $\approx$~4.5~keV, is identified as a proton-CRSF (pCRSF) it would imply either a high-strength ($\sim$10$^{15}$~G) dipole field or a very strong, higher-order multipole field close to the neutron star surface (e.g. Israel et al. 2017b). Alternatively, should the feature be identified as an electron-CRSF (eCRSF) then the dipole (or multipole) field is expected to be far weaker at $\sim$10$^{11}$~G (depending on how far from the neutron star surface the line is actually formed).  

In this letter we present the time-resolved properties of M51 ULX-8 which allows us to rule out the presence of a high strength dipole field and identify the likely nature of the accretion flow.

\begin{figure}
\begin{center}
\includegraphics[trim=100 100 0 0, clip, width=8cm]{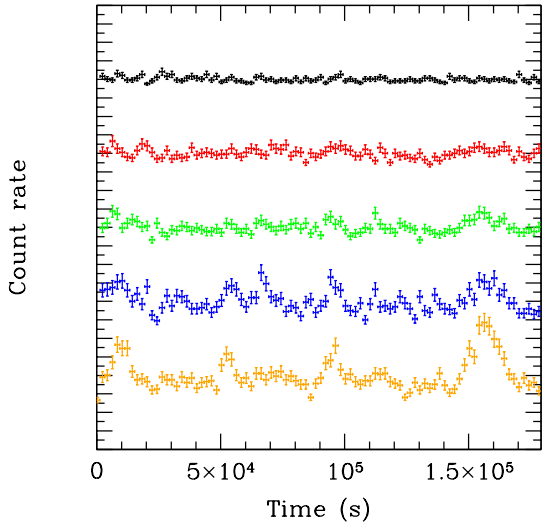}
\end{center}
\vspace{-0.2cm}
\caption{Background-subtracted {\it Chandra} lightcurves of M51 ULX-8 rebinned to 2000s. The energy bands from top to bottom are 0.5--0.7~keV (black), 0.7--1~keV (red), 1--1.3~keV (green), 1.3--2~keV (blue) and 2--8~keV (orange). The lightcurves are offset in count rate for clarity but are on the same absolute scale with small and large ticks indicating 0.005 counts s$^{-1}$ and 0.01 counts s$^{-1}$ respectively. The appearance of long timescale variability at high energies is apparent and may resemble the super-orbital quasi-periodicities seen in other ULPs (but on much shorter timescales in this case).} 
\label{fig:l}
\end{figure}

\section{Data analysis}

Following the procedures outlined in Brightman et al. (2018), we extract and analyse the same {\it Chandra} data of M51 ULX-8 in which the CRSF was discovered --- OBSID: 13813 with an effective exposure of $\approx$~180 ks. In Figure 1 we plot the energy-resolved, background-subtracted lightcurves (0.5--0.7, 0.7--1.0, 1--1.3, 1.3--2 and 2--8~keV) binned on 2000s which indicate that the vast majority of the variability is associated with energies $>$2~keV. Although few in number, we note that the recurrent peaking behaviour at high energies is somewhat reminiscent of the longer timescale behaviour seen in ULPs (e.g. Walton et al. 2016b) but in this case, on far shorter timescales ($\sim$tens of hours rather than $\sim$tens of days in ULPs). Should this behaviour be periodic or quasi-periodic (we note that we have an insufficient number of peaks to confirm any such claim), the timescale would be enormous for the rotation period of the neutron star but {\it could} instead indicate precession of a super-critical accretion flow (e.g. Pasham \& Strohmayer 2013; Middleton et al. 2018). Indeed, the variability does not resemble typical band-limited noise and the clear trend of the source being brighter when spectrally harder (i.e. during the peaks in the hard X-ray band) is a corollary of the latter model (e.g. Middleton et al. 2015).

\begin{figure*}
\begin{center}
\includegraphics[trim= 0 10 10 200, clip, width=18cm]{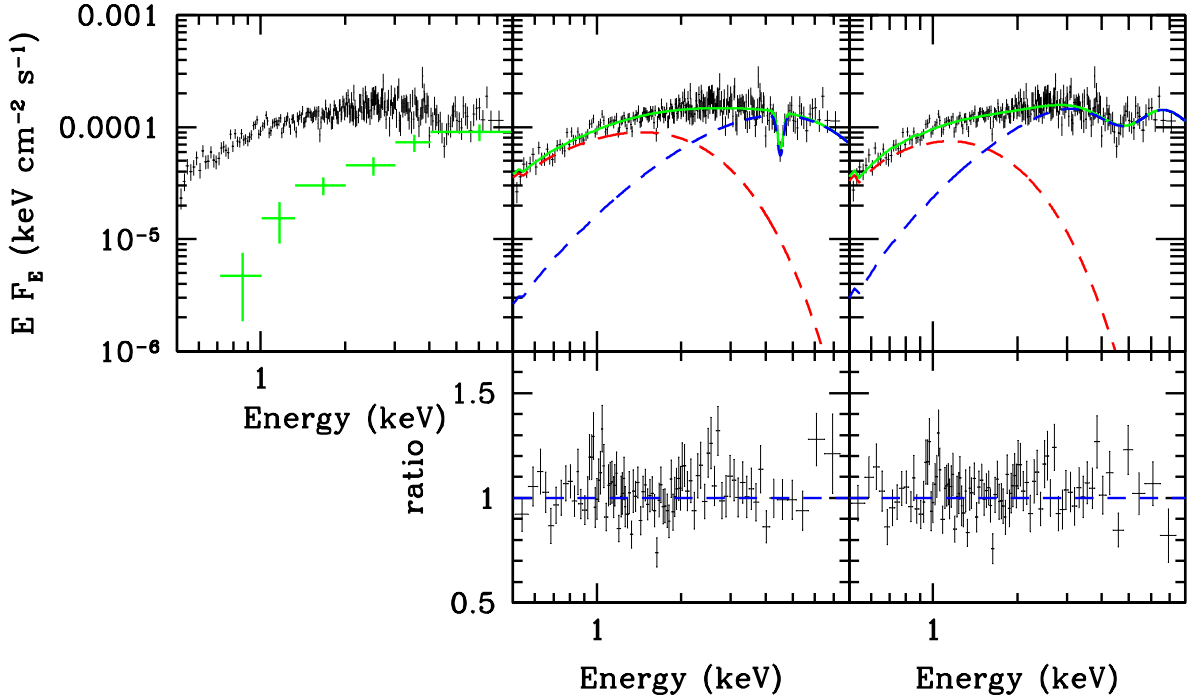}
\end{center}
\vspace{-0.2cm}
\caption{{\it Left}: the covariance (green) and time-averaged (black) spectral data unfolded through a power-law of zero index and unity normalisation (such that there is no bias from unfolding). {\it Centre and right}: The time-averaged data modelled and unfolded through {\sc tbabs*(diskbb + gabs*cutoffpl)} with the parameters of the {\sc cutoffpl} fixed to the best-fitting values from modelling the covariance spectrum (with the exception of the normalisation) and line parameters fixed (centre panel) and free (right-hand panel) respectively. The {\sc diskbb} component is shown in dashed red, the {\sc cutoffpl} component is shown in dashed blue and the combination (including absorption by the CRSF line) is shown in solid green. The lower panels show the residuals to the best-fitting model (data/model) re-binned for clarity in each case.} 
\label{fig:l}
\end{figure*}

We further break the 2-8~keV band into three narrower sub-bands (2--3, 3--4 and 4--8~keV) and obtain an estimate for the power in each and in the pre-existing 0.5--0.7, 0.7--1.0, 1--1.3 and 1.3--2~keV bands by fast-Fourier transforming each lightcurve, broken into 5 segments (with each segment containing 16 $\times$ 2000~s bins). The power in each band is given by $\langle |X_{\nu}|^{2}\rangle = \langle X_{\nu}^{*}X_{\nu} \rangle$ where $X_{\nu}$ is the Fourier transform of the lightcurve segment,  $^{*}$ indicates the complex conjugate and the angle brackets indicate the average over segments (we note that, as the number of segments in this case is only 5, the resulting errors will only be approximately Gaussian). We proceed to obtain the intrinsic (noise-subtracted) coherence (Vaughan \& Nowak 1997) relative to the 1--1.3~keV band from the complex-valued cross spectrum: $C_{\nu} = X_{\nu}^{*}Y_{\nu}$, where $X_{\nu}$ and $Y_{\nu}$ are the Fourier transforms of the reference lightcurve (in this case the 1--1.3~keV lightcurve) and comparison lightcurve respectively at Fourier frequency, $\nu$. The linearly correlated coherence (which measures how well one lightcurve can be mapped to another by a simple linear transformation) is then given by $|\langle C_{\nu}\rangle |^{2}/\langle |X_{\nu}|^{2}\rangle\langle |Y_{\nu}|^{2}\rangle$ where all of the powers in the denominator have been corrected for Poisson noise (and the numerator is corrected following the recipe of Vaughan \& Nowak 1997). From the product of the coherence and noise-subtracted power in the comparison band, we then obtain the covariance (see Wilkinson \& Uttley 2009; Uttley et al. 2014) and integrate from 0.03125--0.25 mHz. 

The covariance in absolute (rms) units is obtained for six out of the seven bins (it cannot be extracted for our lowest energy bin) and is loaded into {\sc xspec} (Arnaud et al. 1996). This covariance spectrum is shown in Figure 2 along with the time-averaged data, both unfolded through the instrumental response, and a power-law of zero index and unity normalisation (such that the process of unfolding does not bias the appearance of the spectrum). The figure indicates that the variable component does not account for the entire 0.3--8~keV emission, requiring at least one separate component at softer energies.

To explore the nature of the source, we require a spectral decomposition which explicitly includes the component revealed by the covariance. We proceed to fit the covariance spectrum with a simple, absorbed, exponentially cut-off power-law ({\sc tbabs*cutoffpl}) with abundances from Wilms et al. (2000) and column density set to that reported in Brightman et al. (2018). The cut-off power-law takes the form: $A(E) = K E^{-\Gamma} \exp(-E/E_{\rm cutoff})$, where $\Gamma$ is the power-law photon index, $E_{\rm cutoff}$ is the e-folding energy of the exponential roll-off and $K$ is the normalisation. This model has previously been used to describe the high energy emission of ULXs (e.g. Bachetti et al. 2013; Pintore et al. 2017) and most recently emission from the accretion column in ULPs (e.g. Walton et al. 2018a, b). Although we do not have the spectral resolution to isolate the contribution (or lack thereof) by the CRSF to the covariance, we include an absorption line (described in {\sc xspec} by a {\sc gabs} component) with centroid energy (4.52~keV) and width (0.11~keV) set by fits to the time-averaged spectrum in Brightman et al. (2018) but with the line normalisation free to vary. A good statistical description of the data is obtained (see Table 1 for parameter values and formal 1$\sigma$ errors) with parameter values ($\Gamma$ and $E_{\rm cutoff}$) for the cut-off power-law similar to those reported for the ULP, NGC 7793 P13 (Walton et al. 2018a). 

We proceed to fit the time-averaged spectrum (background-subtracted and re-binned for chi-squared fitting: see Brightman et al. 2018) with a model composed of a {\sc cutoffpl} component at high energies, a disc blackbody ({\sc diskbb}) at lower energies, and the CRSF ({\sc gabs}) convolved with the {\sc cutoffpl} component only. We fix the index ($\Gamma$) and cutoff energy ($E_{\rm cutoff}$) to the best-fitting values from our fits to the covariance spectrum and allow the normalisation to be free to vary (but fixing the lower limit to the value returned from the fits to the covariance spectrum). We also allow the neutral column to be free to vary, setting the lower limit to the Galactic line-of-sight column density in the direction of M51, 2$\times$10$^{20}$ cm$^{-2}$ (Dickey \& Lockman 1990). We initially fix the CRSF parameters (line energy and width) to those in Brightman et al. (2018) and obtain a statistically acceptable fit to the data (with model parameters and formal 1$\sigma$ errors provided in Table 1 - although we caution that such errors do not necessarily reflect the complex shape of $\chi^{2}$ space in such fits) with the corresponding spectral de-convolution shown in Figure 2. In addition to the model parameters, we extract the unabsorbed 0.3-8~keV integrated flux for the {\sc diskbb} and {\sc cutoffpl} components using the pseudo-model {\sc cflux} and provide the unabsorbed luminosities and errors in Table 1. We note that removing the {\sc diskbb} component and re-fitting the data with only {\sc tbabs*cutoffpl} (with $\Gamma$ and $E_{\rm cutoff}$ still fixed to those from the fit to the covariance spectrum), leads to a increase in $\Delta\chi^{2}$ of $>$3000 for 2 additional d.o.f, confirming that a soft component is statistically required in addition to the cutoff power-law revealed by the covariance.

As the spectral deconvolution differs from that presented in Brightman et al. (2018), we repeat the line significance test they describe, assuming the model is correct and testing for the chance occurrence of a similar or higher $\Delta\chi^{2}$ from adding the {\sc gabs} component at any energy. In this model, the $\Delta\chi^{2}$ from adding the line is -45.14; across 100,000 simulated spectra, only 5 reached this value. Restricting our statistical analysis to only those observations of M51 ULX-8 in which the line was searched for in Brightman et al (2018), we correct the false alarm rate for the additional 8 observations where the line has not been detected (although as Brightman et al. 2018 point out, this may well be expected, as such lines vary and in most cases the data quality cannot rule out the presence of the CRSF). Our final, global false alarm rate is $<$~4$\times$10$^{-4}$ indicating a significance in excess of 3$\sigma$.


We allow the previously fixed CRSF parameters (energy and line-width) to vary and re-fit the time-averaged data. The resulting best-fit invokes a broader line ($\sigma$ $\approx$ 1~keV) with an improvement of $\Delta\chi^{2} \approx$ 5 for 2 d.o.f. Clearly, this is not a statistically significant improvement, yet without a clear requirement to keep the CRSF parameters fixed, we cannot rule out this solution (and indeed, when fitting the covariance spectrum with a broad line instead of a narrow one, neither the fit quality nor model parameters change significantly). Whilst such line parameters might be more in keeping with expectation for an eCRSF ($\sigma/E >$ 0.1 -- Tsygankov et al. 2006), it is apparent from inspection of the residuals to the best-fit that there is significant scatter around the line energy (see Figure 2) and some narrow structure around the previously reported CRSF line centroid. Whilst it is possible that CRSFs in such sources may have complex shapes (e.g. F{\"u}rst et al. 2015) -- perhaps formed of both a broad and narrow component reflecting the dipole and higher-order, multipole fields, density and velocity gradients -- we note that it is highly probable that the broad-line solution is being influenced by the curvature in the continuum. With forthcoming data extending the energy coverage for this source, we will be better able to address this issue. {Due to the obvious issues and uncertainty in this spectral fit we do not repeat the significance test of Brightman et al. (2018) in this case.}

As our bandpass does not extend to the high energies where a hard excess above a Wien tail has been reported for a number of ULXs (e.g. Bachetti et al. 2013; Mukherjee et al. 2015; Walton et al. 2015), we cannot currently determine whether such an excess is present in this source. However, conceivably there could be an additional high energy component whose lower energy emission could extend into our bandpass and impact upon the parameter values we have obtained. To test for this, we introduce a further {\sc cutoffpl} component with parameters based on the fits which isolate the pulsed component in NGC 7793 P13 (Walton et al. 2017), i.e. $\Gamma$ = 0.17, $E_{\rm cutoff}$ = 4.7~keV and free normalisation. We find that, in the case of both of the fits described above, the parameters for the soft component (which will prove crucial in the following analysis) remain unchanged within 3$\sigma$.

\section{testing the dipole field strength}

The spectral fits described above, guided by the covariance, have allowed us to obtain a clearer picture of the true spectral deconvolution for M51 ULX-8. Importantly we now have parameters that can be used to test the picture that the source harbours a strong (10$^{15}$~G) dipole field. This needs to be tested in two distinct cases: where the flow beyond $r_{M}$ is sub-critical or super-critical.

\begin{figure}
\centering
\subfloat[a {\it sub}-critical disc geometry\label{cw_25}]{%
  \includegraphics[trim=200 30 50 0, clip, width=0.45\textwidth]{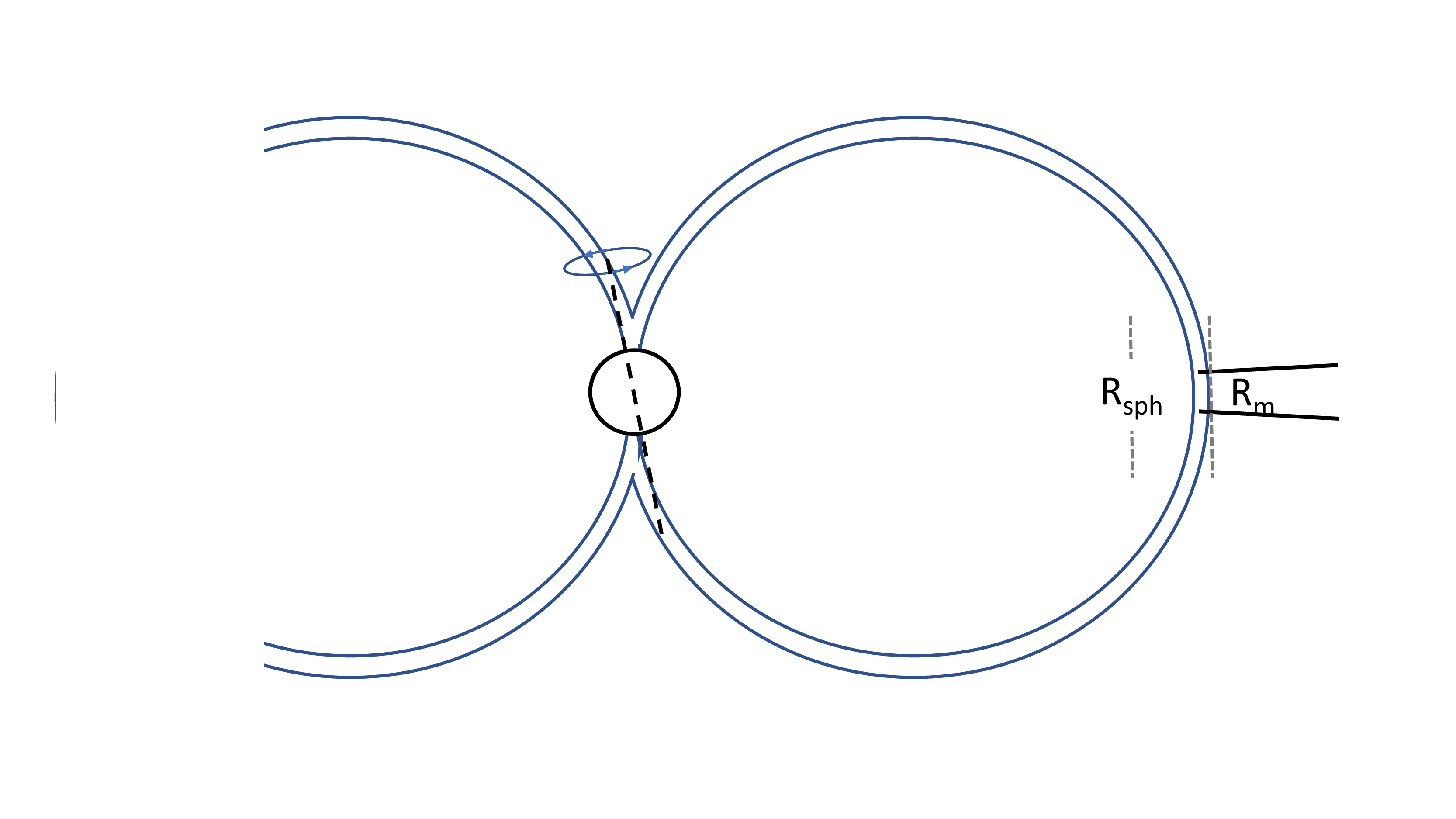}%
  }\par        
\subfloat[a {\it super}-critical disc geometry\label{cw_10}]{%
  \includegraphics[trim=50 30 200 20, clip, width=0.45\textwidth]{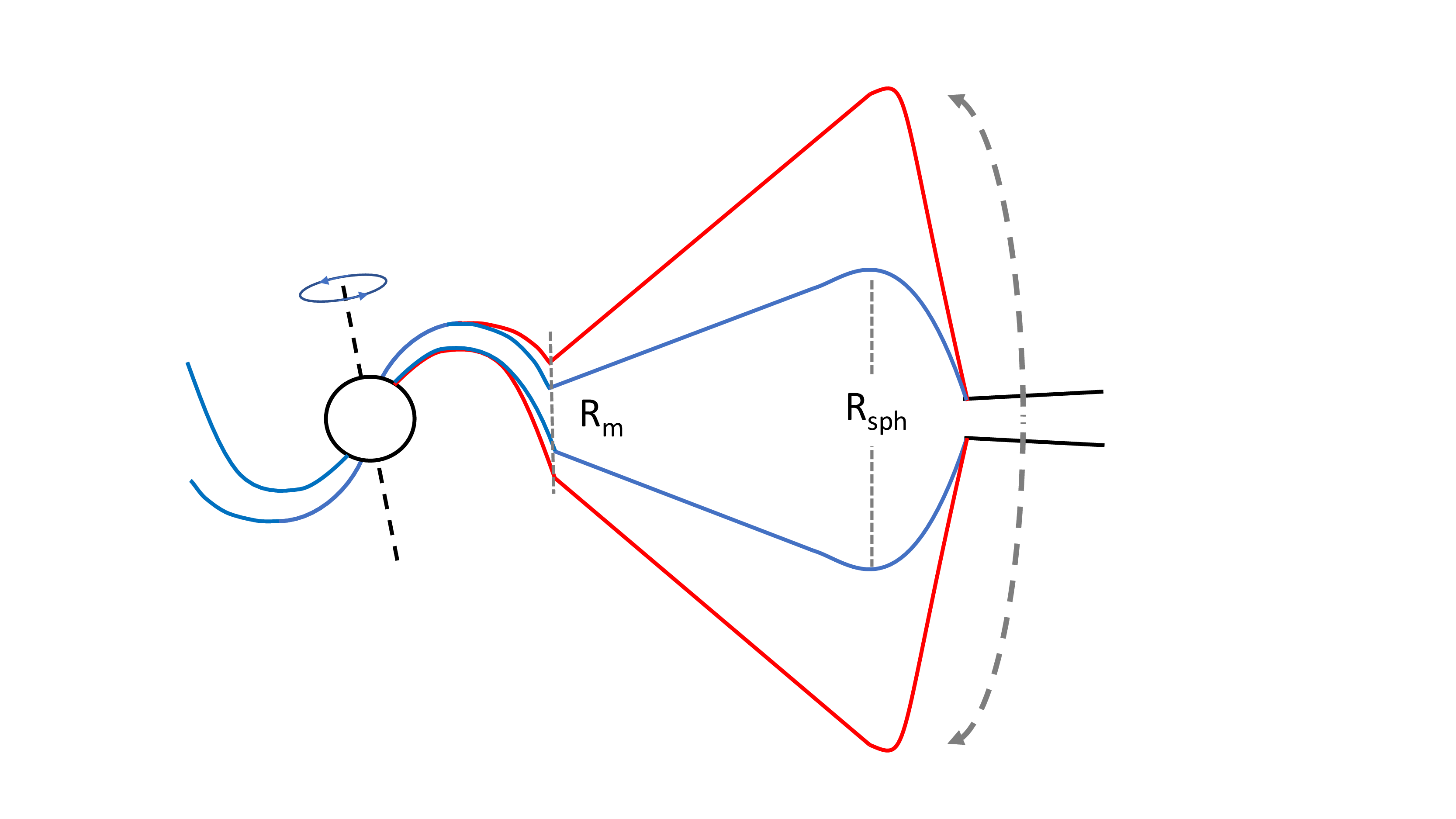}%
  }\par
\caption{A 2D schematic representation of the two models under consideration in the paper. In the top plot we show the {\it sub}-critical disc geometry where the thin disc truncates at $R_{\rm M}$ and forms a closed magnetosphere with an optical depth in this accretion `curtain' determined in part by the magnetic field strength of the neutron star. In the bottom plot we show our favoured model for M51 ULX-8. At large radii the disc is geometrically thin (black lines) but as the Eddington luminosity is reached locally, the disc inflates until radiation pressure supports the flow at $R_{\rm sph}$ (blue lines). Within this radius, mass is lost via winds (red lines). Within $R_{\rm M}$, material is confined to flow along field lines down to the surface of the neutron star where the CRSF is presumably imprinted (and potentially a $\sim$10$^{15}$ G multipole field could be present). Precession of the inflow and outflow (perhaps as a consequence of Lense-Thirring torques, e.g. Middleton et al. 2018) could lead to a changing view of the inner regions as they are partially occulted and produce a correlated variable component at higher energies.}
\label{TS}
\end{figure}

\subsection{A {\it sub}-critical disc}

Depending on mass accretion rate and dipole field strength, it is possible that an accretion disc which would otherwise greatly exceed the Eddington limit onto the neutron star can remain sub-critical (i.e. locally below the Eddington limit). The condition for this to occur is $r_{\rm M} > r_{\rm sph}$, where $r_{\rm sph}$ is the spherisation radius, the point at which the Eddington limit is reached in the disc (see Shakura \& Sunyaev 1973 and Figure 3). Under such a condition, the disc remains geometrically thin and sub-critical down to $r_{\rm M}$ within which the material falls through an accretion curtain onto the neutron star via a shock-heated column (where the decrease in electron scattering cross section can allow for highly super-Eddington rates of infall, e.g Mushtukov et al. 2015). Based on the spectral deconvolution, we would expect the hard, variable component to be from the accretion column whilst the soft component would correspond to the combined emission from the thin disc extending down to $r_{\rm M}$ and accretion curtain (with the latter two having similar spectra when the curtain is optically thick: Mushtukov et al. 2017). 

Should the above scenario be the correct physical description for M51 ULX-8, we require the accretion curtain to be optically {\it thin} such that the CRSF is not suppressed due to repeat scatterings. Mushtukov et al. (2017) determine the scattering conditions in the accretion curtain by considering the dynamics of the gas flowing from $r_{\rm M}$, paying particular note to the geometry of the accretion column (which is affected by the dipole field strength; Mushtukov et al. 2015). The resulting condition for the accretion curtain material to be optically {\it thick} to electron scattering is $L_{\rm 39} \gtrsim B_{\rm 12}^{1/4}$ (where $L_{\rm 39}$ is the accretion luminosity in units of 10$^{39}$ erg~s$^{-1}$ and $B_{\rm 12}$ is the dipole field strength in units of 10$^{12}$ G). Mushtukov et al. (2017) determine that the actual optical depth of the curtain is a function of the distance along the path of the free-falling material and so reversing the above inequality provides only a {\it lower limit} on the field strength required for the entire curtain to be optically thin such that the CRSF is guaranteed to be detected. Given the time-averaged X-ray luminosity of the source in this observation ($\approx$~5$\times$10$^{39}$ erg~s$^{-1}$), we would require field strengths in excess of 6$\times$10$^{14}$~G. It is plausible that a dipole field strength of $\sim$10$^{15}$~G would allow the CRSF to be detected and lead to a somewhat (perhaps only partially) optically thin curtain which would then not leave a substantial imprint on the spectrum due to Comptonisation (although we speculate there should be intrinsic cyclotron emission from the sub-relativistic plasma spiralling along the field lines).

In addition to not suppressing the CRSF, the small number of scatterings of photons from the accretion column will also not significantly diminish any variability from this component (be it intrinsic to the column itself or as a consequence of precession of the dipole: Lipunov \& Shakura 1980; Mushtukov et al. 2017). If we were to assume that the correlated variability we see is indeed from the column, then the lack of variability from the soft X-ray component would be consistent with a geometrically thin disc extending down to $r_{\rm M}$, as the variability from such flows is typically low amplitude and only appears on relatively long timescales dictated by the local viscous frequency (e.g. Lyubarskii 1999; Ingram 2016).

As we will show, there are several issues with this picture for M51 ULX-8. The most obvious is that the soft X-ray component -- which, as we have discussed, might be associated with a thin disc -- has an unabsorbed luminosity at least 8 times the Eddington limit (even for a limiting neutron star mass of 2~M$_{\odot}$: Demorest et al. 2010). In the absence of geometrical beaming (which would invoke a super-critical flow) or some otherwise unknown physical process, this is irreconcilable with a geometrically thin disc. 

Should the spectrum be de-convolved for other ULXs, it is plausible that the soft X-ray component may have sub-Eddington luminosities for a neutron star primary (typically $<$ 2$\times$10$^{38}$ erg s$^{-1}$). Importantly, should a CRSF also be observed, the accretion curtain must be optically thin which may then genuinely allow the soft emission to be identified with a geometrically thin, sub-critical disc. The tests we describe below -- using the parameters of M51 ULX-8 as an example -- take a simple parameter of the spectral model (namely the {\sc diskbb} normalisation) and can be used to determine whether a sub-critical disc is plausible in any given ULX where a neutron star is thought to be present, given the above conditions.

The position of $r_{\rm M}$  is commonly related to the surface dipole field strength and luminosity (e.g. Lamb et al. 1973; Cui 1997; F{\"u}rst et al. 2018) by:

\begin{equation}
r_{\rm M} = 2.7 \times 10^{8}B_{\rm 12}^{4/7}L_{\rm 37}^{-2/7} \hspace{1cm} [\rm cm]
\end{equation}

\noindent which assumes a canonical neutron star (M$_{\rm NS}$ = 1.4 M$_{\odot}$, R$_{\rm NS}$ = 10~km), where $L_{\rm 37}$ is the (bolometric) accretion luminosity in units of 10$^{37}$ erg~s$^{-1}$ (assumed to be a good tracer of the mass accretion rate, with a radiative efficiency $\eta \approx$ 0.21). The normalisation of the {\sc diskbb} component, $N_{\rm dbb}$, is related to the corrected position of the disc inner edge (see Kubota et al. 1998) by:

\begin{equation}
r_{\rm in} = \xi f_{\rm col}^{2}D_{\rm 10}\sqrt\frac{N_{\rm dbb}}{{\rm cos}(\theta)} \hspace{1cm} [\rm km]
\end{equation} 

\noindent where $\xi$ is a correction factor of order unity (determined by Kubota et al. 1998 to be $\approx$ 0.41), $f_{col}$ is the colour temperature correction factor due to scattering/absorption in the disc atmosphere, $D_{\rm 10}$ is the distance to the source in units of 10~kpc (in this case $D$ = 8.58~Mpc --- McQuinn et al. 2016) and $\theta$ is the angle of the disc to the line-of-sight. Combining the above equations (as $r_{\rm in} = r_{\rm M}$ in this sub-critical picture) allows $L_{\rm 37}$ to be estimated for a range in $\theta$ and $B_{\rm 12}$. 


We plot the results in Figure 4 (left-hand plot) for a representative value of $f_{\rm col}$ = 1.7 and using M51 ULX-8 as an example, with the approximate 1$\sigma$ range in $N_{\rm dbb}$ from our spectral modelling, of 0.1 to 1.22, found by allowing the CRSF parameters to be fixed or free respectively (see Table 1). 

In Figure 4 (right-hand plot), we also plot the predicted luminosity as a function of dipole field strength at the limit of the sub-critical disc model -- the point where $r_{\rm M} = r_{\rm sph}$. We use the approximation, $r_{\rm sph} \approx \dot{m}_{\rm 0}r_{\rm isco}$ (Shakura \& Sunyaev 1973), where $\dot{m}_{\rm 0}$ is the mass transfer rate through the outer disc in units of the Eddington mass accretion rate at $r_{\rm isco}$ (the ISCO radius which we assume to be at 6GM/c$^{2}$). As the disc is still (just) sub-critical in this picture, we assume that $L \propto \dot{m}_{\rm 0}$ (rather than in a super-critical disc where $L \approx L_{\rm Edd}[1+ln(\dot{m}_{\rm 0})]$ - Shakura \& Sunyaev 1973) such that:

\begin{figure*}
          \centering
\includegraphics[trim=80 0 0 0, clip, width=9cm]{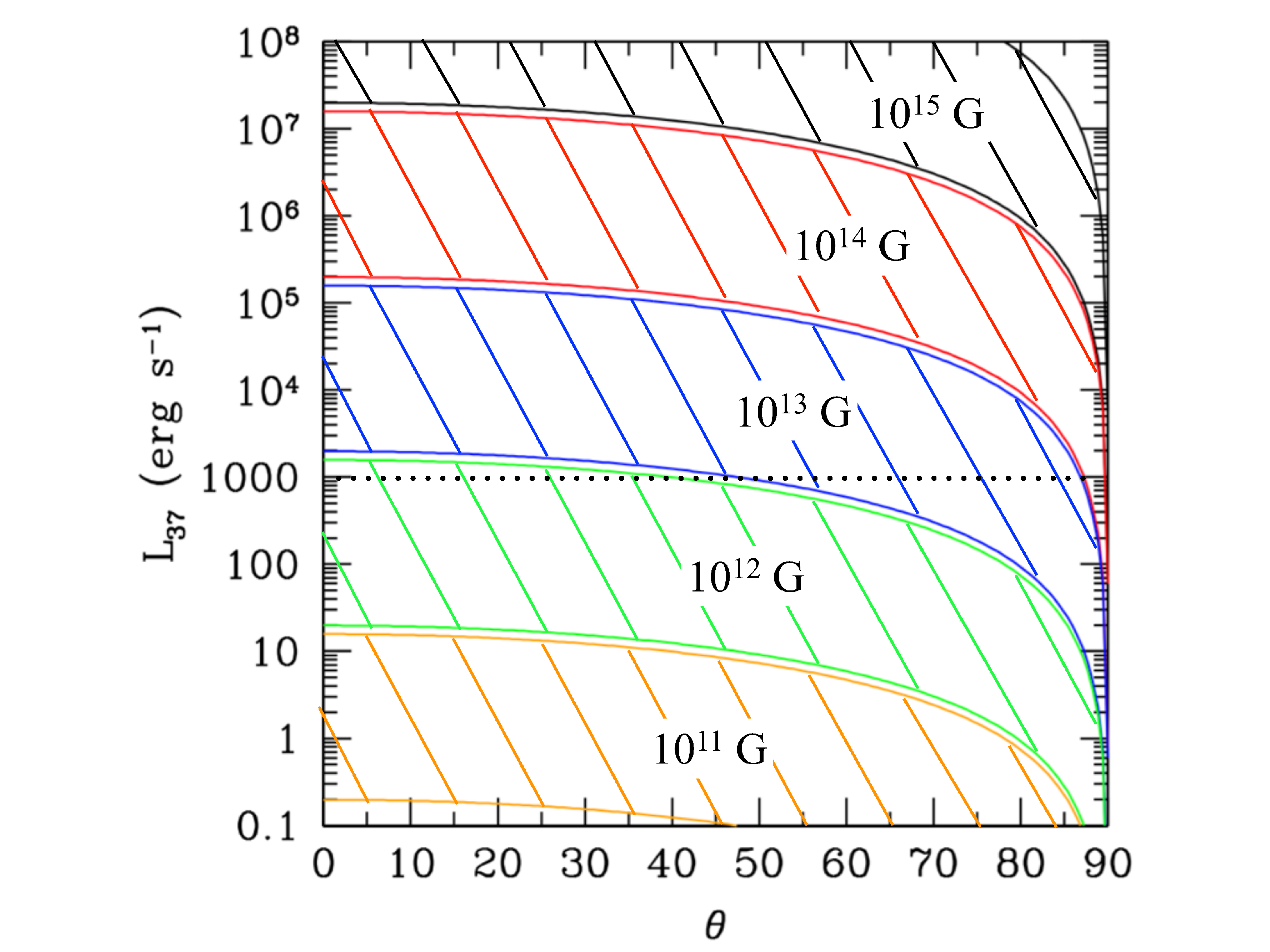}%
  \includegraphics[trim=20 47 0 0, clip, width=10.3cm]{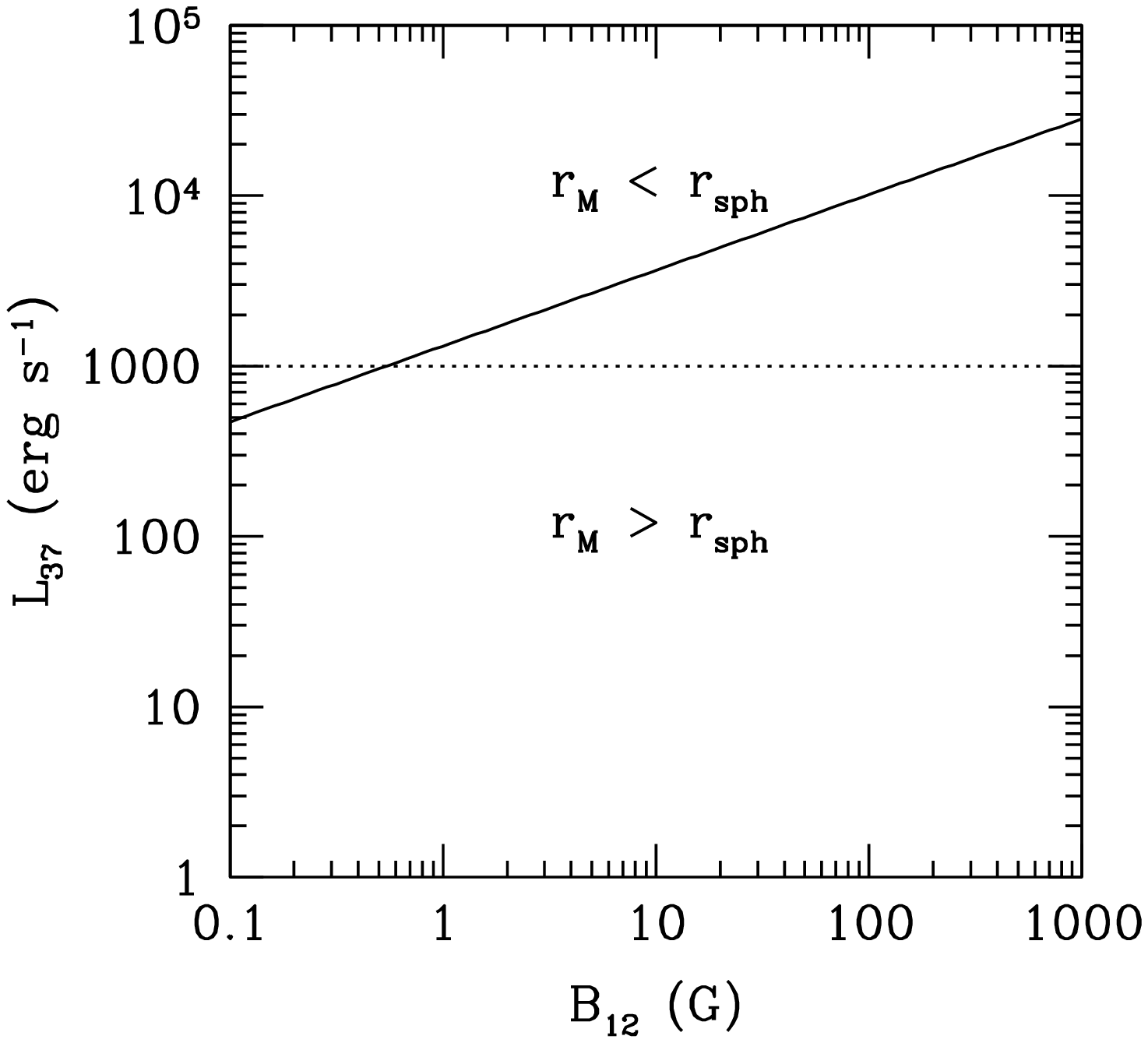}
  \vspace {0.5cm} 
  \caption{An example of the tests that can be applied to ULXs considered to harbour neutron stars where the soft component can be associated with a geometrically thin disc (e.g. when a CRSF is present such that the accretion curtain is optically thin). Left: the implied luminosity of M51 ULX-8 (in units of 10$^{37}$ erg~s$^{-1}$) versus inclination, from the combination of equations (1) and (2) (i.e. a model which assumes a disc where the Eddington limit is not reached locally) where $r_{\rm M} = r_{\rm in}$ and the {\sc diskbb} normalisation is taken from Table 1. The range of values for each dipole field strength results from the putative maximum range in $N_{\rm dbb}$ (assuming either case presented in Table 1) whilst the horizontal dotted line refers to the approximate observational limit on the luminosity from this source of 1$\times$10$^{40}$ erg~s$^{-1}$. Right: the limiting luminosity for the sub-critical model (once again in units of 10$^{37}$ erg~s$^{-1}$) versus dipole field strength (in units of 10$^{12}$ G) which occurs when $r_{\rm M} = r_{\rm sph}$. From the left-hand plot, it is clear that the sub-critical model -- based on the parameters for M51 ULX-8 --  becomes invalid at field strengths in excess of $\sim$10$^{13}$ G as the inferred luminosity from the {\sc diskbb} normalisation would place the source into the super-critical regime in the right-hand plot.} 
       \end{figure*}

\begin{equation}
\dot{m}_{\rm 0} = \frac{\dot{m}}{\dot{m}_{\rm Edd}} = \frac{0.08 L_{37}}{0.21 L_{\rm Edd,37}} \approx 0.03L_{\rm 37}m_{\rm NS}^{-1}
\end{equation} 

\noindent where $m_{\rm NS}$ is the neutron star mass in M$_{\odot}$ which we take to be 1.4, 0.08 is the radiative efficiency at the ISCO (which underpins the formula for $r_{\rm sph}$ although noting that, in reality the radiative efficiency at the ISCO is probably closer to 0.06 for neutron star systems - Sunyaev \& Shakura 1986). Combined with equation (1), we then obtain the general formula $L_{\rm 37} \approx 1300 B_{\rm 12}^{4/9}$ for $r_{\rm M} = r_{\rm sph}$.

As can be seen from inspection of Figure 4 (right-hand plot), in order to maintain a sub-critical disc with a dipole field of 10$^{15}$ G implies luminosities {\it not} in excess of $\approx $10$^{41}$ erg~s$^{-1}$. Whilst this is consistent with observations of M51 ULX-8, it is several orders of magnitude {\it below} that which would be inferred from our use of the {\sc diskbb} parameters for the same dipole field strength (Figure 4, left-hand plot). Although it is possible to hide much of the luminosity should the source be (almost exactly) edge-on, this is both highly unlikely and, in the case of M51 ULX-8, no eclipses have been reported over several observations (three of which are in excess of a day in duration - Brightman et al. 2018). In keeping with our initial expectation based on the inferred luminosity, we have found that the model for sub-critical accretion onto a neutron star with a dipole field strength of 10$^{15}$~G is indeed invalid in the case of M51 ULX-8.

We note that Figure 4 would appear to suggest that lower dipole field strengths ($<$10$^{12}$G) {\it are} consistent with producing both the observed total luminosity and still be consistent with the above requirement to remain sub-critical (with the notable exception of the disc luminosity). However, a key requirement is that the accretion curtain be optically thin, which must be the case for M51 ULX-8 as the CRSF is observable; for a time-averaged luminosity of 5 $\times$ 10$^{39}$ erg~s$^{-1}$ this is not possible for dipole field strengths below $\approx$ 6 $\times$ 10$^{14}$G (Mushtukov et al. 2017). 

As a further test of the validity of a geometrically thin disc in the spectrum of ULXs, we can consider whether it can avoid being radiation pressure dominated at $r_{\rm M}$  (noting that in the case of M51 ULX-8 we know that this cannot be the case given the luminosity is well in excess of the Eddington value for even a very massive neutron star). The disc is expected to become radiation pressure dominated when (Padmanabhan 2001; Frank et al. 2002; Andersson et al. 2005): 


\begin{equation}
r \lesssim 880 \alpha^{2/21} \dot{m}_{\rm 0}^{16/21}\left(\frac{m_{\rm NS}}{1.4}\right)^{1/3} [\rm km]
\end{equation} 


\noindent where $\alpha$ is the viscosity coefficient of Shakura \& Sunyaev (1973), we have assumed that the opacity is dominated by electron scattering and have made the explicit assumption that this radius in the disc is much greater than that of the neutron star (see Andersson et al. 2005). Substituting for (3) and equating to (2) then leads to the condition that the disc is {\it not} radiation pressure dominated when:

\begin{equation}
N_{\rm dbb} \gtrsim \frac{1.8\times10^{4}}{f_{\rm col}^{4}D_{\rm 10}^{2}} \alpha^{4/21}L_{\rm 37}^{32/21}m_{\rm NS}^{-6/7} {\rm cos}(\theta)
\end{equation} 

\noindent 
Assuming $\alpha$ = 0.1, $f_{\rm col}$ = 1.7 and $L_{\rm 37} \approx$ 500, we find that, to obtain the range of normalisations from our fits, would imply that the disc could only avoid being radiation dominated for $\theta \gtrsim$ 86\degree which, as we have argued previously, is highly unlikely. 












\subsection{A {\it super}-critical disc}

We have strong evidence that, in the case of M51 ULX-8, a sub-critical disc cannot be accommodated and we must instead explore the case for a super-critical flow, i.e. $r_{\rm M} < r_{\rm sph}$ for a range in dipole field strengths. The key difference compared to the previous physical scenario is that, from $r_{\rm sph}$ down to $r_{\rm M}$, the disc is now geometrically thick, with advection and mass loss via winds stabilising the accretion flow. Whilst the super-Eddington luminosity in the sub-critical case had to rely on a magnetic pressure supported accretion column and suppression of the electron scattering cross section in the presence of a strong magnetic field, here we have both a luminosity from a radiatively supported disc ($L \approx L_{\rm Edd}[1+{\rm ln}(\dot{m}_{\rm 0})]$) and collimation of emission from within $r_{\rm sph}$ (which leads to a breakdown in the assumption of isotropy and results in geometrical beaming, e.g. King 2009; Middleton \& King 2016). In the case of a sub-critical flow, the soft X-ray component can be associated with a thin disc (providing it meets the criteria discussed in this paper), whereas in the super-critical model, the soft X-ray component is assumed to be associated with emission from the spherisation radius where the disc is instead geometrically thick (see Figure 3). Such a geometry and the relevant temperatures in the spectra are presented in Poutanen et al. (2007); although this model was developed assuming a black hole primary, to first order we would not expect the presence of a neutron star primary to make a substantial difference to the soft X-ray emission -- in the absence of direct irradiation by the neutron star, $r_{\rm sph}$ is independent of compact object type. Conversely - and deviating somewhat from the black hole model of Poutanen et al. (2007), the harder X-ray emission must be associated with some combination of the inner disc, accretion curtain and column (see for instance the simulations of Takahashi et al. 2018; Abarca et al. 2018).

Based on the formulae of Poutanen et al. (2007), the temperature of the soft component found from our spectral modelling implies $\dot{m}_{\rm 0} \approx$ 20 (for $f_{\rm col}$ = 1.7). In order to establish whether this is compatible with a 10$^{15}$~G dipole field, we determine the critical accretion rate at which $r_{\rm M}$ = $r_{\rm sph}$. In units of the Eddington rate this is $\dot{m}_{\rm 0} \approx 28 B_{\rm 12}^{4/9}$ (or in commonly used units of 10$^{17}$ g~s$^{-1}$ as $\dot{m}_{\rm 17} \approx 700 B_{\rm 12}^{4/9}$ where we have assumed the radiative efficiency at the ISCO is $\approx$ 0.08 - although see Sunyaev \& Shakura 1986). From this, we determine that, for a 10$^{15}$~G dipole field, $r_{\rm M}$ = $r_{\rm sph}$ for $\dot{m}_{\rm 0} \approx$~600, far above that implied by associating the soft component with emission from $r_{\rm sph}$. As $\dot{m}_{\rm 0} \approx$ 20 with B =10$^{15}$ G would result in $r_{\rm M} > r_{\rm sph}$, we appear to be unable to construct a working model, be it sub- or super-critical, for M51 ULX-8 where the dipole field strength is so high.

Our spectral analysis and the restrictions described above, allow us to investigate whether a dipole field, similar in strength to those typically found in Galactic HMXBs (e.g. F{\"u}rst et al. 2014; Tendulkar et al. 2014; Yamamoto et al. 2014; Bellm et al. 2014) and potentially other ULPs at $\lesssim$10$^{13}$ G (e.g. King \& Lasota 2016; King, Lasota \& Kluzniak 2017; Christodoulou et al. 2016; F{\"u}rst et al. 2016), could account for the spectrum, luminosity and still allow for the observation of a CRSF. Using the previous formula, we determine that, for $\dot{m}_{\rm 0} \approx$ 20, we expect an upper limit on the dipole field strength of $\sim$10$^{12}$ G and the flow to be super-critical (with $r_{\rm sph}\approx 120 R_{\rm g}$). As $r_{\rm M} < r_{\rm sph}$ in this case, the luminosity of the source results from the combination of the super-critical flow (minus that lost in an outflow), that from the accretion column (assuming B $> 10^{9}$ G) and geometrical beaming. Whilst, in the sub-critical case such low field strengths would imply the accretion curtain is optically thick (Mushtukov et al. 2017), geometrical beaming of the emission by a factor of only a few allows the accretion curtain to start to become optically thin (see earlier discussion), thereby allowing us to detect the CRSF (when the view to the inner regions is not totally obscured by the thick disc and winds).

\section{Discussion \& Conclusion}

The degeneracy often inherent in time-averaged spectral fitting can be broken through time-resolved approaches, as we have demonstrated in the case of M51 ULX-8. Interpreting the $\approx$~4.5~keV line feature as a pCRSF implies either a dipole field of $\sim$10$^{15}$~G or a higher-order field of this strength closer to the neutron star surface (and falling off more rapidly with distance). As pointed out in Brightman et al. (2018), the width of the CRSF can indicate the responsible species, with electron CRSFs being typically broader ($\sigma/E_{cyc} \sim$ 0.1: Tsygankov et al. 2006) than their proton counterparts ($\sigma/E_{cyc} <$ 0.1: Ibrahim et al. 2002). Our analysis indicates that a broad component to the line cannot yet be ruled out although this may be a consequence of the limited energy coverage, and narrow structure is almost certainly still present (Figure 2).

Irrespective of the true shape of the line (and contributions from either the proton or electron populations), our use of the covariance has allowed us to effectively rule out a 10$^{15}$~G dipole field in this source; in such a scenario, the luminosity is inconsistent with the sub-critical interpretation and the accretion rate required to be super-critical is at odds with the spectrum. However, we stress that we do not rule out a $\sim$10$^{15}$~G multipole field in this source. 

We determine the likely upper bound for the dipole field strength to be $\sim$10$^{12}$ G (consistent with the field strengths estimated from the spin-up rates seen in ULPs: e.g. King \& Lasota 2016; Christodoulou et al. 2016; F{\"u}rst et al. 2016;  King, Lasota \& Kluzniak 2017; Carpano et al. 2018) and find that only a super-critical flow is able to reproduce the luminosity, spectrum and optically thin accretion curtain (the latter allowing us to observe the CRSF). In such a flow, the disc has a large scale-height from $r_{\rm sph}$ down to $r_{\rm M}$ and precession of this flow could potentially result in the repeated changes in flux seen in the hard spectral component from the inner-most regions (e.g. Dauser, Middleton \& Wilms 2017); such a geometry is presented in Figure 3. 

Should a multipole field (assumed to be $\sim$10$^{15}$~G) be present in addition to the $<$~10$^{12}$ G dipole, a requirement of our model is that it should not dominate over the latter at the magnetospheric radius. This in turn requires that the field fall off as $R^{-n}$ where $n \gtrsim$ 3 [{\rm log}($R_{\rm NS}/R_{\rm M}$)-1]/{\rm log}($R_{\rm NS}/R_{\rm M}$). In the case of our inferred super-critical disc with $\dot{m}_{\rm 0}$ = 20, this implies $n \gtrsim$ 5 (at the limit of $r_{\rm M} = r_{\rm sph}$). Although this would appear to preclude the presence of a quadrupole field, we note that we have assumed the formulae relating to super-critical discs (in determining the accretion rate and location of $r_{\rm sph}$) are free of uncertainties. Although unlikely to be the case, should they be accurate, we still cannot rule out an even higher order multipole field (e.g. a hexapole or octopole; P{\'e}tri 2015) leading to the reported CRSF.



A corollary of associating a more classical super-critical accretion flow to M51 ULX-8 is that we expect to detect line-of-sight atomic features in absorption (the mass-loaded, radiatively driven wind) and isotropic emission lines (perhaps due to shock heating between the wind and secondary star) as seen in other ULXs (Middleton et a. 2014, 2015b; Pinto et al. 2016, 2017; Walton et al. 2016a) and most recently in a ULP (Kosec et al. 2018). Detection of such atomic features in the spectrum of M51 ULX-8 in future (perhaps with the advent of {\it $\chi$RISM} or {\it Athena}) would provide further confirmation regarding the nature of the inflow in this source. Whilst the CRSF in the spectrum of M51 ULX-8 is the clearest example of such a signal found to-date in a ULX, discovering more CRSFs directly or indirectly (e.g. Walton et al. 2018c) should allow the approaches we have outlined in this paper to be applied to other sources suspected of harbouring neutron stars (assuming the continuum can be deconvolved) and the nature of the accretion flow to be firmly established.
 
\section{Acknowledgements}

The authors thank the anonymous referee. MJM and DJW appreciate support from Ernest Rutherford STFC fellowships.

\label{lastpage}

\end{document}